\documentclass{aastex631} 

\newcommand\rrl{RRL}
\newcommand\rrls{RRLs}

\defcitealias{2021ApJ...908...20C}{C21}
\defcitealias{Mullen2021}{Paper I}
\defcitealias{Mullen2022}{Paper II}

\received{November 14, 2022}
\revised{January 9, 2023}
\accepted{January 9, 2023}

\shorttitle{Gaia DR3 PLZ}
\shortauthors{Mullen et al.}
\graphicspath{{./}{figures/}}

\begin{document}

\title{RR~Lyr\ae{} mid-infrared Period-Luminosity-Metallicity and Period-Wesenheit-Metallicity relations based on Gaia DR3 parallaxes}

\correspondingauthor{Joseph P. Mullen}
\email{jpmullen@iastate.edu}

\author[0000-0002-1650-2764]{Joseph P. Mullen}
\affiliation{Department of Physics and Astronomy, Iowa State University, Ames, IA 50011, USA}

\author[0000-0001-9910-9230]{Massimo Marengo}
\affiliation{Department of Physics, Florida State University, 77 Chieftain Way, Tallahassee, FL 32306, USA}
\affiliation{Department of Physics and Astronomy, Iowa State University, Ames, IA 50011, USA}

\author[0000-0002-9144-7726]{Clara E. Mart\'inez-V\'azquez}
\affiliation{Gemini Observatory, NSF's NOIRLab, 670 N. A'ohoku Place, Hilo, HI 96720, USA}
\affiliation{Cerro Tololo Inter--American Observatory, NSF's NOIRLab, Casilla 603, La Serena, Chile}

\author[0000-0003-3096-4161]{Brian Chaboyer}
\affiliation{Department of Physics and Astronomy, Dartmouth College, 6127 Wilder Laboratory, Hanover, NH 03755, USA}

\author[0000-0002-4896-8841]{Giuseppe Bono}
\affiliation{Dipartimento di Fisica, Universit\`a di Roma Tor Vergata, via della Ricerca Scientifica 1, 00133 Roma, Italy}
\affiliation{INAF -- Osservatorio Astronomico di Roma, via Frascati 33, 00078 Monte Porzio Catone, Italy}

\author[0000-0001-7511-2830]{Vittorio F. Braga}
\affiliation{IAC- Instituto de Astrof\'isica de Canarias Calle V\'ia Lactea s/n, E-38205 La Laguna, Tenerife, Spain\\}
\affiliation{INAF -- Osservatorio Astronomico di Roma, via Frascati 33, 00078 Monte Porzio Catone, Italy}

\author[0000-0001-8209-0449]{Massimo Dall'Ora}
\affiliation{INAF -- Osservatorio Astronomico di Capodimonte, Salita Moiariello 16, 80131 Napoli, Italy}

\author[0000-0002-2662-3762]{Valentina D'Orazi}
\affiliation{Dipartimento di Fisica, Universit\`a di Roma Tor Vergata, via della Ricerca Scientifica 1, 00133 Roma, Italy}
\affiliation{INAF -- Osservatorio Astronomico di Padova, Vicolo dell’Osservatorio 5, 35122, Padova, Italy}

\author[0000-0001-5829-111X]{Michele Fabrizio}
\affiliation{INAF -- Osservatorio Astronomico di Roma, via Frascati 33, 00078 Monte Porzio Catone, Italy}
\affiliation{Space Science Data Center -- ASI, via del Politecnico snc, 00133 Roma, Italy}

\author[0000-0001-5292-6380]{Matteo Monelli}
\affiliation{IAC- Instituto de Astrof\'isica de Canarias Calle V\'ia Lactea s/n, E-38205 La Laguna, Tenerife, Spain\\}
\affiliation{Departmento de Astrof\'isica, Universidad de La Laguna, E-38206 La Laguna, Tenerife, Spain\\}

\author[0000-0002-5032-2476]{Fr\'ed\'eric Th\'evenin}
\affiliation{Universit\'e de Nice Sophia--antipolis, CNRS, Observatoire de la C\^ote d'Azur, Laboratoire Lagrange, BP 4229, F-06304 Nice, France}




\begin{abstract}
We present new empirical infrared Period-Luminosity-Metallicity (PLZ) and Period-Wesenheit-Metallicity (PWZ) relations for RR~Lyr\ae{} based on the latest Gaia EDR3 parallaxes. The relations are provided in the WISE $W1$ and $W2$ bands, as well as in the $W(W1, V - W1)$ and $W(W2, V - W2)$ Wesenheit magnitudes. The relations are calibrated using a very large sample of Galactic halo field RR~Lyr\ae{} stars with homogeneous spectroscopic [Fe/H] abundances (over 1,000 stars in the $W1$ band), covering a broad range of metallicities ($-2.5 \la \textrm{[Fe/H]}\la 0.0$). We test the performance of our PLZ and PWZ relations by determining the distance moduli of both galactic and extragalactic stellar associations: the Sculptor dwarf spheroidal galaxy in the Local Group (finding $\bar{\mu}_{0}=19.47 \pm 0.06$), the Galactic globular clusters M4 ($\bar{\mu}_{0}=11.16 \pm 0.05$) and the Reticulum globular cluster in the Large Magellanic Cloud ($\bar{\mu}_{0}=18.23 \pm 0.06$). The distance moduli determined through all our relations are internally self-consistent (within $\lesssim$ 0.05 mag) but are systematically smaller (by $\sim$ 2-3$\sigma$) than previous literature measurements taken from a variety of methods/anchors. However, a comparison with similar recent RR~Lyr\ae{} empirical relations anchored with EDR3 likewise shows to varying extents a systematically smaller distance modulus for PLZ/PWZ RR~Lyr\ae{} relations.
\end{abstract}



\section{Introduction} \label{sec:intro}

RR Lyrae stars (RRLs) have been extensively utilized as a primary standard candle within our Local Group of galaxies. They are easily identifiable, with a period between 0.2 and 1 day, and are found nearly everywhere due to being evolved $\sim$0.7 solar mass stars. As such, RRLs are the most widely used tracers of old (age $\ga$ 10 Gyr, \citealt{Walker1989,2020A&A...641A..96S}) stellar populations in the local neighborhood (see, e.g., \citealt{2016MmSAI..87..358B} and \citealt{2015pust.book.....C} for a review). Besides their use in probing galactic structures, an accurate calibration of RRL distances allows for a cosmological distance ladder completely based on Population II stars, independent from Classical Cepheids. This has assumed great relevance over the last decade due to the current tension in measured values of the Hubble constant, which is still mainly anchored on classical Cepheids (see \citealt{2022arXiv220801045R}). 

RRLs were first used as standard candles due to their evolutionary position on the Horizontal Branch (HB) yielding a nearly constant $V$-band magnitude, with a minor metallicity dependence \citep{1958RA......5....3B}. This correlation ultimately leads to a well-defined $M_{V}$ vs. iron abundance relation \citep{1990ApJ...350..603S, 1998A&ARv...9...33C}; however, in its simplicity, this relation overlooks evolutionary effects on the HB and is prone to errors in accounting for extinction and possible non-linearities \citep{2003MNRAS.344.1097B}. While brightness in the V-band does not show any significant dependence on the period, Period-Luminosity-Metallicity relations (PLZ) do exist in the infrared (see e.g. \citet{2015ApJ...808...50M, 2017ApJ...841...84N} for theoretical relations, and \citet{2013ApJ...776..135M,2013MNRAS.435.3206D, 2018MNRAS.481.1195M, 2019MNRAS.490.4254N,2021MNRAS.503.4719G} for observational ones). These PLZ relations have an accuracy approaching other traditional stellar standard candles that can be characterized by a Leavitt Law \citep{1908AnHar..60...87L,1912HarCi.173....1L}, such as Cepheids. The accuracy improves for the \rrl{} PLZ relationships when moving to the infrared as both the importance of accounting for reddening decreases and the dispersion of the PLZ itself decreases, due in part to the dependence on metallicity increasing and the pulsational amplitude decreasing with longer wavelengths. Moving to the infrared minimizes temperature effects, as infrared RRL observations are mostly determined by the radius variation during the star's pulsation rather than the effective temperature changes that dominate in the optical wavelengths \citep{2016MmSAI..87..358B}. We can similarly combine optical bands to create a Wesenheit magnitude that by construction is reddening independent, as well as less susceptible to temperature effects. These magnitudes can be effectively used in Period-Wesenheit-Metallicity relations (PWZ), in lieu of an infrared PLZ \citep{2015ApJ...808...50M}.

Until recently, PLZ and PWZ relations were limited in their applications due to the scarcity of accurate high-resolution (HR, R $\ga$ 20,000) spectroscopic metallicity measurements, capable of providing accuracy of $\sim$ 0.1 dex in [Fe/H] for individual RRL. Indeed, these relations have been predominantly based on RRL residing in globular clusters (GCs) with well-studied cluster metallicity. The age of large area photometric time surveys (e.g., ASAS-SN \citep{2014ApJ...788...48S, 2018MNRAS.477.3145J}, Catalina Sky Survey \citep{2009ApJ...696..870D}, PanSTARRS \citep{2016arXiv161205560C}, DES \citep{2018ApJS..239...18A}, Gaia \citep{2016A&A...595A.133C,2019A&A...622A..60C}, TESS \citep{2015JATIS...1a4003R}, ZTF \citep{2019PASP..131a8002B}, OGLE \citep{2005AcA....55...43S}), have, however, heralded a new age where large samples of RRL variables, especially in the field of the Galactic Halo and Bulge, are finally available. Using effectively these new RRL catalogs, however, still requires knowing their metallicities, which can be prohibitive if it requires taking spectra of such a large number of targets. This issue can be avoided if the metallic abundance for these stars could be  estimated by studying the properties of their light curves, without the need of collecting their spectra. With this motivation in mind, in \citealt{Mullen2021} (hereafter \citetalias{Mullen2021}) and \citealt{Mullen2022} (hereafter \citetalias{Mullen2022}) we have provided new relations to derive photometric metallicities of fundamental and overtone RRL stars respectively, based on their optical $V$-band and infrared (WISE W1 and W2 bands, \citealt{2010AJ....140.1868W}) light curves. The distinction of our relations from other available photometric metallicities is that they are based on the largest available sample of homogeneous spectroscopic metallicities available to date (over 9,000 \rrls{}), which we published in \citet{2021ApJ...908...20C} (\citetalias{2021ApJ...908...20C} hereafter) and in \citet{2021ApJ...919..118F}. In this paper we expand on our previous work by providing PLZ and PWZ relations based on the same sample of calibrating RRLs. Metallicities for this sample are still derived from \citetalias{2021ApJ...908...20C}. Distances are instead obtained from the most recent Gaia EDR3 parallax values \citep{2022arXiv220800211G}. At the time of this publication, there have already been some Gaia EDR3 distance calibrations using either smaller calibration samples or different bands/relations than those presented in this work (e.g., \citealt{2021MNRAS.503.4719G,2022MNRAS.513..788G,https://doi.org/10.48550/arxiv.2206.07668}). Overall, the relations presented in this work are based on the largest available calibrating sample with individual spectroscopic metallicities obtained on a homogeneous scale, and as such they allow for studying the dependence of PLZ and PWZs over the entire [Fe/H] abundance range in the Galaxy.

This paper is structured as follows. In Section~\ref{sec:dataset}, we describe the data sets we adopt for our work: the metallicity catalog utilizing the work of \citetalias{2021ApJ...908...20C}, derivation of mean magnitudes from optical and infrared time-series, and the Gaia parallax measurements. In Section~\ref{subsec:calibration}, we explain how our PLZ/PWZ relations are obtained and calibrated. Section~\ref{subsec:application} assesses the precision of the Wesenheit and infrared relations as we apply our method to derive the distance modulus to M4 (a Galactic globular cluster), Reticulum (a globular cluster in the Large Magellanic Cloud), and Sculptor (a Local Group dwarf spheroidal galaxy) and compare our distance moduli with previous literature measurements. Our conclusions are presented in Section~\ref{sec:conclusion}.

\section{Datasets} \label{sec:dataset}

Our calibration sample of RRLs is chosen from an extensive catalog of 8660 field RRLs, from which we have either [Fe/H] abundances derived from HR spectra ($R \ga 20$,000) or an estimate of their metallicity based on the $\Delta$S method \citep{1959ApJ...130..507P}. This HR+$\Delta$S metallicity catalog is comprised of both the high-resolution metallicity catalog of \citetalias{2021ApJ...908...20C} and the application of the spectrum selection criteria and $\Delta$S metallicity calibration of \citetalias{2021ApJ...908...20C} (applied in \citealt{2021ApJ...919..118F}) to the full medium-resolution ($R \sim 2$,000) Large Scale Area Multi-Object Spectroscopic Telescope (LAMOST) DR2 survey \citep{2012RAA....12..735D, 2014IAUS..298..310L} and Sloan Extension for Galactic Understanding and Exploration \citep[SEGUE,][]{2009AJ....137.4377Y} datasets. By utilizing metallicity measurements from the single work of \citetalias{2021ApJ...908...20C}, we ensure our entire sample is on a homogeneous metallicity scale (consistent with the globular cluster metallicity scale provided by \citealt{Carretta2009}). For a complete and detailed description of the metallicity scale, the HR metallicity catalog's demographics, the $\Delta$S calibration, and the spectrum selection criteria, we refer the reader to the \citetalias{2021ApJ...908...20C} paper. 

The resultant HR+$\Delta$S metallicity catalog was cross-matched with the $Gaia$ EDR3 database \citep{EDR3, EDR3cat,EDR3pi} to provide astrometric data. Finally, photometry is obtained by cross-matching the variables in our metallicity catalog with both the Wide-field Infrared Survey Explorer (WISE, \citealt{2010AJ....140.1868W}), and its Near-Earth Objects reactivation mission (NEOWISE, \citealt{2011ApJ...731...53M}) in the 3.4 and 4.6 $\mu$m bands, $W1$, and $W2$ respectively. Similarly, visible time-series photometric measurements in the $V$ band were taken from the All-Sky Automated Survey for Supernovae (ASAS-SN, \citealt{2014ApJ...788...48S, 2018MNRAS.477.3145J}). For a more detailed description of the NEOWISE, ASAS-SN survey and how we have integrated such measurements to our calibration sample, we refer the reader to  \citetalias{Mullen2021} and \citetalias{Mullen2022}.

To compute new PLZ relationships, periods and mean magnitudes of this sample were derived by following the procedure from \citetalias{Mullen2021}, summarized below. The period of each RRL star was refined using the Lomb-Scargle method \citep{1976Ap&SS..39..447L, 1982ApJ...263..835S} applied to the ASAS-SN and WISE time series data. The surveys' large temporal baseline ($\ga$ 8 years) allows us to determine well-defined periods with average accuracy on the order of $\sim 10^{-6}$ days, which properly phase the data without any readily detectable shifting in phase. Mean magnitudes for each band were determined by applying a Gaussian locally-weighted regression smoothing (GLOESS) algorithm to the phased data to gain a smoothed light curve. Each \rrl{} star's mean magnitude is determined by integrating the smoothed light curve flux over one period, then subsequently converting back into magnitude. For more specifics on the GLOESS algorithm and the calculation of mean magnitude from a GLOESS light curve with its uncertainty, we refer you to \citet{2004AJ....128.2239P} or \citet{2015ApJ...808...11N}, respectively.

To ensure a clean sample of \rrls{}, a number of quality cuts were made on the input catalog. We first implement the various quality checks on light curve quality and data extraction as elaborated upon in \citetalias{Mullen2021}. Quality checks include: comparing refined period to that of literature, defining the maximum allowable scatter around the GLOESS light curve, and implementing iterative rounds of outlier rejection for individual epochs relative to the GLOESS phased lightcurve. Furthermore, in order to eliminate possible contact eclipsing binary (W UMa) from our dataset, whose sinusoidal light curves can often be confused with the light curves of first overtone \rrl{}, we required all the \rrl{} in our sample to have photometric measurements in both $V$ and $W1$. Following the procedure of \citet{Mullen2022} (Section 3.2), the amplitude ratio between the $V$ and $W1$ magnitudes serves as a primary discriminator between \rrl{} stars and eclipsing binaries. Note that this criterion does not cause any issues with sample size as the ASAS-SN survey has nearly complete coverage ($\sim 98\%$) of the WISE stars used in this work. Subsequent quality cuts based upon astrometric criterion or due to the fitting method utilized are noted in the following section. The exact number of stars used in each calibration, after outlier rejection, is given in Table~\ref{tab_fits}.

\section{Calibration of PLZ and PWZ relations\label{subsec:calibration}}

Gaia EDR3 parallaxes were processed according to the recommendations in \citet{2022arXiv220800211G}. In particular, the parallaxes were corrected for the $Gaia$ zero-point systematic error using the calibration of \cite{EDR3bias}.  Guided by \S 7.1.2 of the EDR3 documentation and Figure 19 in \cite{EDR3cat}, the uncertainty in the EDR3 parallaxes have been increased from their catalog values by $\sim 10\%$ to $60\%$ depending on the magnitude of the star and its solution type in EDR3.

Reddenings were determined from the \citet{Stilism} 3-D maps. If the parallax uncertainty was less than 10\% of the parallax, the distance to each star was estimated from its parallax, otherwise, the distance, as a first order approximation, was calculated from the PLZ relation of \cite{2021MNRAS.503.4719G} in the $W1$ band (assuming no reddening). After the initial reddening determination, the distance was recalculated and the reddening determined again. A comparison of the two reddening determinations indicated that two iterations was sufficient to lead to convergence in the reddening determination. Conversion from E(B-V) to extinction assumed $R_V = 3.1$ from \citet{1989ApJ...345..245C}, and  $A_{W1}/A_V = 0.061$,  $A_{W2}/A_V = 0.048$ from the \citet{2013MNRAS.430.2188Y} extinctions. In the W1 sample, 93\% of the stars have relatively low reddenings, with  $E(B-V) < 0.2$, which corresponds to $A_{W1} < 0.038\,$mag.  In the W2 sample, 89\% of the stars have $E(B-V) < 0.2$. The maximum reddening in the sample is $E(B-V) = 0.67$, corresponding to $A_{W1} = 0.13\,$mag. Thus, uncertainties in the reddening values, considered to be $\sim$10\%, will not significantly impact the PLZ fit. 

In order to obtain a clean sample of \rrls{}, a number of additional quality cuts were made on the input catalog. Stars which may have spurious astrometric solutions can be identified by a large value of the renormalised uniform weight error \citep[RUWE,][]{EDR3pi}, and so only stars with $RUWE < 1.4$ were used in the fitting process ($\sim 10\%$ of \rrl{} were removed due to this). A plot of the \texttt{astrometric\_excess\_noise} as a function of RUWE revealed that there were some stars which had large excess astrometric noise which had RUWE $< 1.4$. As a result, an additional cut of  \texttt{astrometric\_excess\_noise} $< 0.2$ ($\sim 2\%$ of the data) was applied to the catalog prior to fitting the $PLZ$ relation.  

\begin{figure*}
\begin{center}
    \includegraphics[scale=0.8]{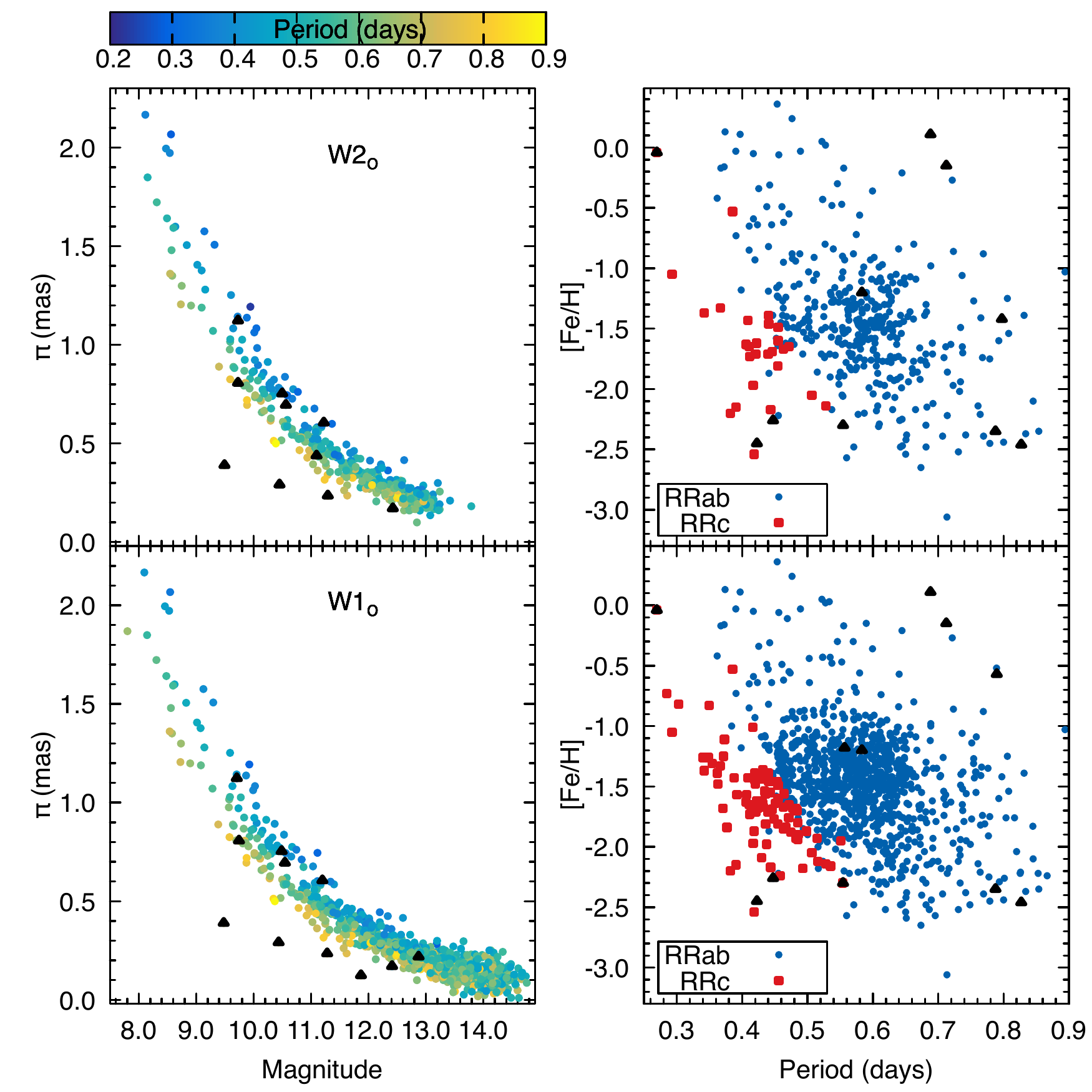}
\end{center}
 \caption{The photometric and parallax properties of the star used in the PLZ fits (left panel), along with their Period--[Fe/H] distributions (right panel). The top two panels show the W2 sample, while the bottom two panels show the W1 sample.  Magnitudes have been dereddened, and RRc stars have had their periods fundamentalized by adding $0.127$ to $\log\,P_{FO}$. The black triangles are the stars which are more than $4\,\sigma$ outliers and which were discarded from the final fit.
 \label{fig_data} }
 \end{figure*}

The overall properties of the data sample used to determine the PLZ relation are shown in Figure \ref{fig_data}. There are over 1000 stars in the W1 sample (9\% with HR spectroscopic [Fe/H]) and about 400 stars in the $W2$ sample (22\% with HR [Fe/H]); Table~\ref{tab_fits} contains the exact sample size used for each relation. The existence of a period-luminosity (PL) relationship is clearly visible in the left panel of Figure \ref{fig_data}, as RRLs with shorter periods are intrinsically less luminous than longer-period stars. Thus, at a given apparent magnitude, the stars with a shorter period have to be closer and hence have a larger parallax than stars with longer periods. The sample covers a broad [Fe/H] range, from $-2.5$ to 0.0 dex. Although their uncertainties are not shown in this figure, the parallaxes are of high quality ($\bar{\sigma}\approx0.02$ mas) -- there are only 48 stars in the W1 sample (all with $W1 > 13.45\,$mag) which have $\varpi/\sigma_\varpi < 2$, while the lowest quality parallax in the W2 sample has $\varpi/\sigma_\varpi = 2.5$.

\begin{figure*}
\begin{center}
  \includegraphics[scale=0.8]{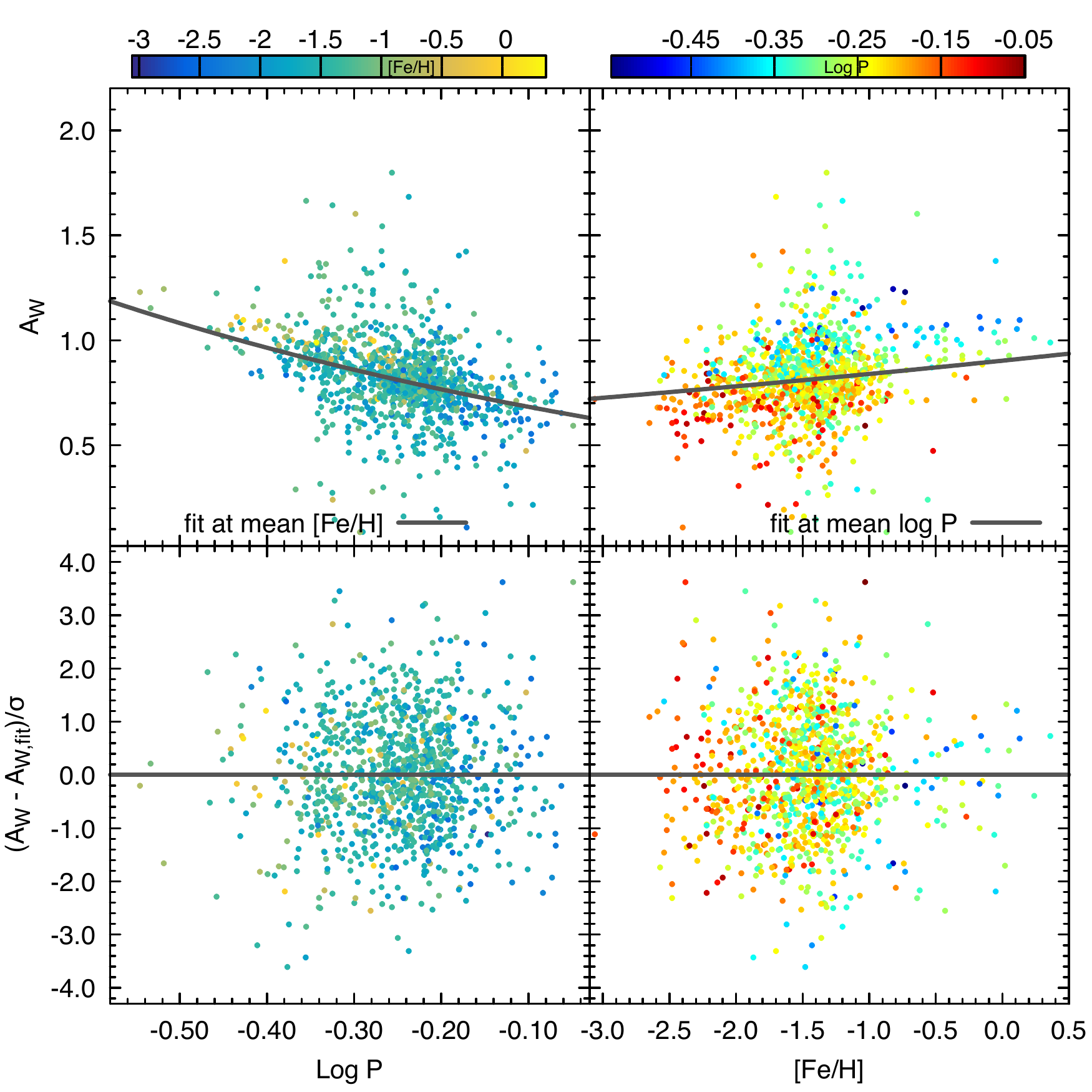}
\end{center} 
 \caption{The P$W1$Z fit after the rejection of $\>4\,\sigma$ outliers. The vertical axis in the top panel is the ABL magnitude computed in the $W1$ band, $A_{W1} = \varpi 10^{0.2 m_{W1,o} - 2}$. Uncertainties $A_{W1}$ (top panels) are not shown, but can be inferred by looking at the normalized fit residuals shown in the bottom panels. Points are colored by their [Fe/H] values in the left panels, and their $\log P$ values in the right panels. 
 \label{fig_W1fit} }
 \end{figure*}

The PLZ calibration largely followed the procedure outlined in \cite{2021MNRAS.503.4719G} and \cite{Layden19}. In brief, the PLZ was fit using the Astrometric Based Luminosity (ABL) $A_{m_o}$, defined as:
 \begin{equation}
\varpi 10^{0.2m_o - 2} = 10^{0.2 [a (\log P + 0.27) + b ([\mathrm{Fe/H}] + 1.3) + c ] }
 \label{eqnfitfunctionE}
\end{equation}
 where $\varpi$ is the parallax in mas, $m_o$ is the absorption corrected apparent magnitude, and the $a,b$ and $c$ are determined in the fit.  The RRc stars had their period fundamentalized by adding 0.127 to $\log P$ \citep{1971A&A....14..293I,2022MNRAS.tmp.2792B}.  The explicit, non-linear fit, which takes into account the uncertainties in all of the observed quantities was performed using a weighted orthogonal distance regression \citep{Boggs, Zwolak} and the nonlinear fitting function \texttt{nls} in R \citep{rstats}.
 An intrinsic dispersion in the PLZ at constant $\log P$ and [Fe/H] of  $0.02\,$mag to $0.04\,$mag was assumed to exist in the PLZ relation during the fit process, consistent with the intrinsic dispersion found in mono-metallicity clusters such as M4 and Reticulum. The initial PLZ fit indicated that there were some outliers ($>4\,\sigma$ residuals from the fit) in the data. These $>4\,\sigma$ outliers led to a poor fit (as judged by a quantile-quantile plot of the fit residuals) and were removed from the fit, before the final fit was performed. As the number of outliers was never large (at most 11 stars, which is less than 1\% of the W1 sample, and less than 3\% of the W2 sample) this outlier rejection did not substantially change the fit coefficients, but did lead to a much higher fit quality and a reduced uncertainty in the fit coefficients. The results of this fit for the $W1$ filter are illustrated in Figure \ref{fig_W1fit}. Fits for the W2 filter look similar, though with $\sim 1/2$ the number of points.
 
\begin{figure*}
\begin{center}
  \includegraphics[scale=0.8]{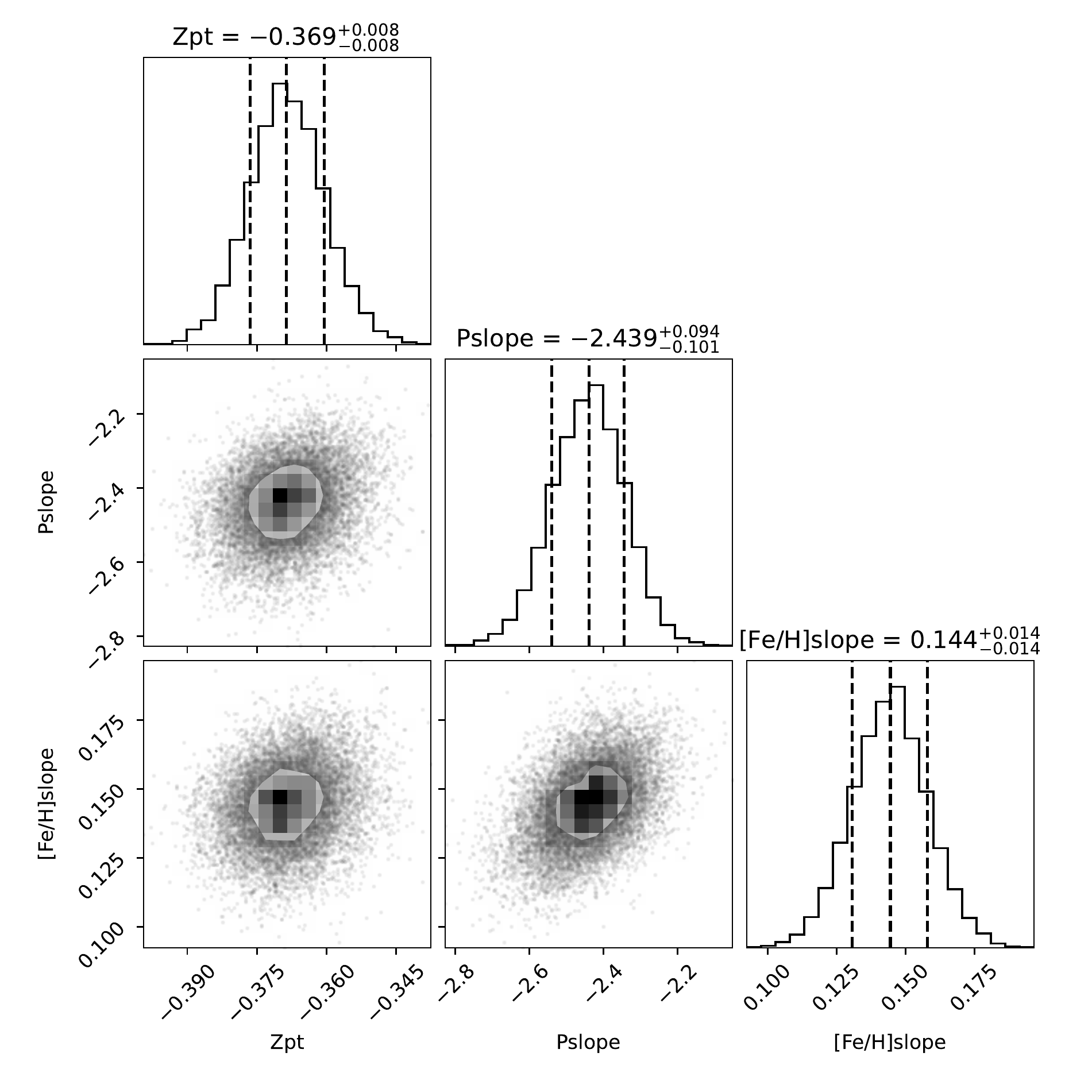}
\end{center} 
 \caption{The distribution of the P$W1$[Fe/H] fit coefficients determined by bootstrap resampling. In the figure `Zpt' stands for zero-point (coefficient $c$ in Table \ref{tab_fits}); `Pslope' stands for the log Period slope (coefficient $a$) and the slope with [Fe/H] (coefficient $b$) is called `[Fe/H]slope'.
 \label{fig_W1corner} }
 \end{figure*}

In order to accurately estimate the uncertainties in the fitted coefficients, bootstrap resampling was used to generate 10 000 data sets, and the fit was performed 10 000 times. The bootstrap resampled datasets have the same number of points as the original dataset (see Table~\ref{tab_fits}), but the points in the dataset was selected randomly (with replacement) from the original dataset. When picking the points for the bootstrap sample, the value of a particular datapoint was selected from the uncertainty distribution (assuming Gaussian uncertainties in [Fe/H], $\varpi$, period, magnitude and reddening) associated with that datapoint. The results of this fitted process are summarized in Table \ref{tab_fits}. The distribution of the fit coefficients for the $W1$ fit is shown in Figure \ref{fig_W1corner}. As expected, given the correlation between period and [Fe/H] in \rrls{}, there is a strong correlation between the log period slope and the [Fe/H] slope.  

In addition to fitting $PLZ$ relations in the $W1$ and $W2$ filters, absolute luminosity relations were also found for the reddening free Wesenheit magnitudes $W(W1,V) = W1 - (V-W1)*0.061/  ( 1 - 0.061)$ and 
$W(W2,V) = W2 - (V-W2)*0.048/ ( 1 - 0.048)$, and these fit coefficients are reported on the third and fourth line of Table \ref{tab_fits}. As a check to determine the sensitivity of these results to the assumed dispersion in the $PLZ$ (at constant period and [Fe/H]) and the astrometric quality cut RUWE, the $W(W1,V)$ fit was performed for a few different values of these parameters (RUWE$<1.4$ with $\sigma_{disp}=0.04$ and RUWE$<1.2$ with $\sigma_{disp}=0.04$). The trial fitted coefficients were within $1\,\sigma$ of their value from the first fits; therefore, the reported fit uncertainties provide a reasonable estimate of the true uncertainties in the PLZ fits. 

\begin{table*}
 \centering
 \caption{PLZ and PWZ relationship coefficients defined as $M=a(\log P + 0.27) +b([Fe/H]+1.3)+c$. The number of stars in a given fit is denoted by $N$, the assumed intrinsic dispersion in the PLZ relation is denoted by $\sigma_{disp}$ $\alpha$ is the color coefficient used in the Wesenheit magnitude.
 \label{tab_fits}}
 \begin{tabular}{lccccccc}
 Magnitude & $N$  & RUWE & $\sigma_{disp}$  & $\alpha$\tablenotemark{a} & $a$ & $b$ & $c$ \\
 \hline
 W1 & 1052 & $<1.4$ & 0.02 & \nodata & $-2.44\pm 0.10$ & $0.144\pm 0.014$ & $-0.369\pm0.008$ \\
 W2 & 397 &  $<1.4$& 0.02 & \nodata & $-2.54\pm 0.10$ & $0.151 \pm 0.014$ & $-0.367 \pm 0.009$ \\
 W(W1, V-W1) & 1054 & $<1.4$&  0.02 & 0.065 & $-2.55\pm 0.10$ & $0.13 \pm  0.015$ & $-0.438 \pm 0.008$  \\
 W(W2, V-W2)& 399 & $<1.4$ &  0.02 & 0.050 & $-2.62\pm 0.10$ & $0.14 \pm 0.015$ & $-0.419 \pm 0.009$ \\
  \hline
 \end{tabular}
 \end{table*}

\section{Distances of three stellar associations in the Milky Way and beyond \label{subsec:application}}

We selected three old stellar systems with multi-epoch \rrl{} photometry available in $V, W1$, and $W2$ bands in order to test the PLZ and PWZ relations listed in Table~\ref{tab_fits} by measuring their distance moduli: Messier~4 (M4, NGC~6121) and Reticulum (two globular clusters, the first Galactic, and the second associated to the Large Magellanic Cloud), and the Sculptor dwarf spheroidal (dSph) galaxy in the Local Group. The three systems are between metal poor and metal intermediate, with $\rm{[Fe/H]} = -1.66$ dex \citep{2004MNRAS.352..153M} for Reticulum, $\rm{[Fe/H]} = -1.10$ dex \citep[see][]{2015ApJ...799..165B} for M4, and $\rm{[Fe/H]} = -1.85$ dex \citep{Mullen2022} for Sculptor dSph. The three systems contain numerous known \rrls{}: 32 in Reticulum  \citep{1976ApJ...208..932D, 1992AJ....103.1166W}, 47 in M4  \citep{2001AJ....122.2587C,2014PASP..126..521S}, and 536 in Sculptor dSph \citep{10.1093/mnrasl/slw093}.

For this analysis, we used the mean $V$ magnitudes published by \citet{2013AJ....145..160K} for Reticulum, by \citet{2014PASP..126..521S} for M4, and by \citet{10.1093/mnrasl/slw093} in Sculptor dSph. The infrared photometry available for each system is in the Spitzer IRAC 3.6 and 4.5 $\mu$m filters \citep[][respectively]{2018MNRAS.480.4138M, 2015ApJ...808...11N, 2018MNRAS.481..578G}, similar to $W1$ and $W2$, and showing no significant offset in mean magnitude as shown by \citet{2017ApJ...841...84N}. It is worth noting that only $W1$ photometry was available for Sculptor dSph, not $W2$. 

For Reticulum and M4 we have derived four independent distance moduli: two of them using the $W1$ and $W2$ PLZ relations, and two more using the PWZ relations based on the $W1$ and $W2$ bands, combined with the $V$-band in the visible. For Sculptor dSph we only derived two distance moduli based on the $W1$ PLZ, and the $V$-$W1$ PWZ. For this reason we have limited out \rrl{} samples to the stars that, in each system, have photometry in both optical and infrared bands. In addition to that, we have applied some quality criteria (described below) intended to remove outliers in the distance measurement.

For Reticulum, we followed the same methodology as in \citet{2021MNRAS.503.4719G} and reject from the 30 RRL stars with $V, W1$, and $W2$ measurements,  those stars that \citet{2018MNRAS.480.4138M} does not include in their calculations of the PL (i.e., V01, V08, V19, V24, V28, V32) since their position on the color-magnitude diagram is unusual or they have noisy light curves, leaving 24 \rrls{} in our final sample. In the case of M4, following \citet{2015ApJ...808...11N}, we reject two stars (V20 and V21) for being blended with nearby sources. Thus, we end up with 31 RRL stars having $W1$ mean magnitudes and 28 with $W2$ mean magnitudes. For Sculptor dSph, we only select the \rrl{} classified by \citet{10.1093/mnrasl/slw093} containing both $V$ and $W1$ photometry, ending up with 42 stars.

In order to get the true distance modulus ($\mu_0$) from the PLZ relationships, we dereddened the $W1$ and $W2$ photometry. We adopted $E(B-V) = 0.03$ \citep{1992AJ....103.1166W} for Reticulum, $E(B-V) = 0.37$ \citep{2012AJ....144...25H} for M4, and $E(B-V) = 0.018$ \citep{2008AJ....135.1993P} for Sculptor dSph. The reddening was then converted into extinction in the WISE bands using the relations $A_{W1} = 0.203~E(B-V)$ and $A_{W2} = 0.156~E(B-V)$ \citep{2012ApJ...759..146M} for Sculptor dSph and Reticulum (assuming $R_V = 3.1$). For M4, the ratio of total to selective absorption is different ($R_V = 3.62$, \citealt{2012AJ....144...25H}), so we used the following extinction values, $A_{W1} = 0.251~E(B-V)$ and $A_{W2} = 0.193~E(B-V)$. We used these latter extinction coefficients to update M4's $\alpha^*$ coefficients for the Wesenheit magnitudes, with $\alpha^*(W1,V-W1) = 0.075$ and $\alpha^* (W2,V-W2) = 0.056$ \citep{2021MNRAS.503.4719G}.

The distance moduli we obtained for each system are listed in Table~\ref{tab:calibrators}. The quoted uncertainties are the sum of the systematic and random uncertainties of each measurement. The former are obtained by propagation of errors considering the photometric uncertainties of the mean magnitudes in $V$, $W1$, and $W2$, the uncertainties of the coefficients in the relationships (see Table~\ref{tab_fits}), and the nominal uncertainties of 0.2 dex in [Fe/H] and $10^{-4}$ days in the period. In the case of the distances calculated from PLZ, we also take account of the uncertainty that comes from the reddening value, considered to be 10\%. The random uncertainties are instead estimated from the standard error of the mean, i.e., the standard deviation divided by the square root of the number of RRL stars used to derive the distance modulus.

\begin{table}[!ht]

    \begin{center}
    \caption{Distance Moduli}\label{tab:calibrators}
    \begin{tabular}{llc}
    \tableline
    \tableline
    System & Filter & $\mu_{0}$ [mag] \\
    \tableline
    Reticulum & PW(W1,V-W1) & 18.22$\pm$0.10 \\
    Reticulum & PW(W2,V-W2) & 18.23$\pm$0.11 \\
    Reticulum & W1 & 18.23$\pm$0.10 \\
    Reticulum & W2 & 18.24$\pm$0.12 \\
    \tableline
    M4 (NGC 6121) & PW(W1,V-W1) & 11.19$\pm$0.08 \\
    M4 (NGC 6121) & PW(W2,V-W2) & 11.14$\pm$0.08 \\
    M4 (NGC 6121) & W1 & 11.18$\pm$0.08 \\
    M4 (NGC 6121) & W2 & 11.13$\pm$0.08 \\
    \tableline
    Sculptor dSph & PW(W1,V-W1) & 19.46$\pm$0.08 \\
    Sculptor dSph & W1 & 19.47$\pm$0.08 \\
    \tableline
    \end{tabular}  
    \end{center}
\end{table}

\begin{figure*}
\centering
\includegraphics[scale=0.5]{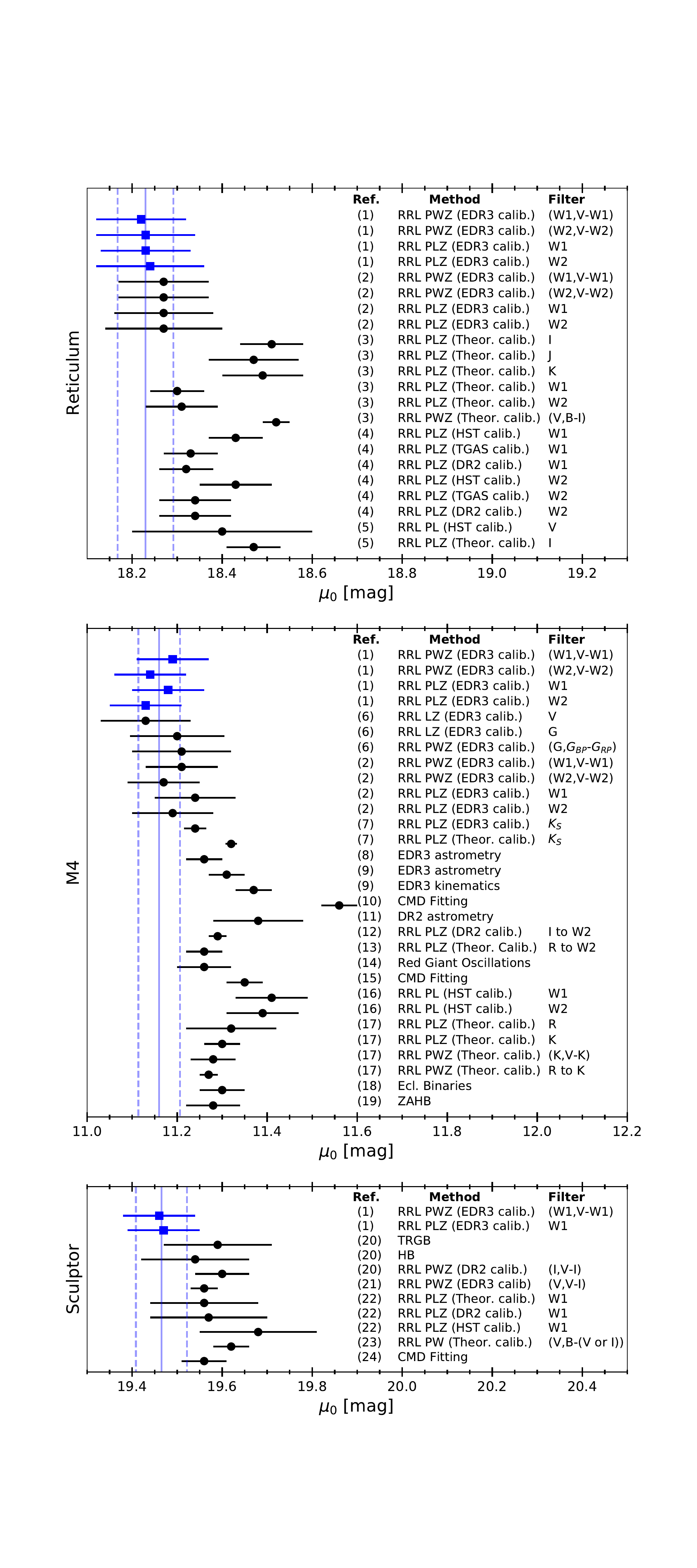}
\vspace{-2cm}
\caption{Comparison of the true distance moduli from this work (blue squares) and literature from the last decade (black circles).
The vertical lines represent the mean and statistical error (see text) of this work's measurements. Beside each measurement is listed a number corresponding to the reference used to make this figure (below), method used to estimate the distance modulus, and observation band. \textit{\textbf{References}} -
(1) This work;
(2) \citealt{2021MNRAS.503.4719G};
(3) \citealt{Braga_2019};
(4) \citealt{2018MNRAS.480.4138M};
(5) \citealt{2013AJ....145..160K};
(6) \citealt{2022MNRAS.513..788G};
(7) \citealt{2021ApJ...909..200B};
(8) \citealt{Vasiliev_2021};
(9) \citealt{2021MNRAS.505.5957B};
(10) \citealt{Valcin_2020};
(11) \citealt{2019MNRAS.489.3093S};
(12) \citealt{2019MNRAS.490.4254N};
(13) \citealt{2017ApJ...841...84N};
(14) \citealt{2016MNRAS.461..760M};
(15) \citealt{2016ApJ...823...18C}
(16) \citealt{2015ApJ...808...11N};
(17) \citealt{2015ApJ...799..165B};
(18) \citealt{2013AJ....145...43K};
(19) \citealt{2012AJ....144...25H};
(20) \citealt{2022arXiv220508548T};
(21) \citealt{Nagarajan};
(22) \citealt{10.1093/mnras/sty2222};
(23) \citealt{2015MNRAS.454.1509M};
(24) \citealt{2014ApJ...789..147W}.}
 \label{fig:moduli}
 \end{figure*}

Figure \ref{fig:moduli} compares the distance moduli calculated through this work's PLZ/PWZ relations (blue squares) to a selection of literature distance measurements from the last ten years (i.e., from 2012) for these three stellar systems. Next to each measurement, we list the method used to estimate the distance modulus; for methods based upon RRL, we also note within parentheses how the zero-point of the calibration was obtained. The error bars plotted are those given by the original study, which for some works listed are just the random errors, not accounting for additional uncertainties such as zero-point calibration, and therefore may be underestimated. It is worth noting that the different works are based upon different methods and/or stellar calibrators. They possess different intrinsic dispersion, and sometimes they may be based on fixed assumptions such as reddening and metallicity in the case of PLZ/PWZ. We refer the reader to explore the individual references listed for full details about the calibration utilized and determined distance moduli.  

Overall, the distance moduli estimated with this work's various PLZ/PWZ relations are in good agreement with each other for Reticulum (within 0.02 mag) and Sculptor dSph (within 0.01 mag) and quite similar for M4 (within 0.05 mag), with an average distance modulus (blue vertical lines in Figure~\ref{fig:moduli}) of $\bar{\mu}_{0}=18.23 \pm 0.06$ for Reticulum, $\bar{\mu}_{0}=11.16 \pm 0.05$ for M4, and $\bar{\mu}_{0}=19.47 \pm 0.06$ for Sculptor dSph. The errors noted on the average distance moduli are just the statistical error due to photometry, i.e., the uncertainty divided by the square root of the number (N) of independent bands used in the average (N=2 for Sculptor and N=3 for M4 and Reticulum). An analysis of Figure~\ref{fig:moduli} shows that our distance moduli and those RRL relations based on an EDR3 anchor tend to be consistently smaller than other distance moduli determinations in the literature. For Reticulum, we note that the distance modulus is smaller than the LMC ($\mu_0\approx$18.4 mag); however, some discrepancy is expected, as Reticulum is widely separated from the center of the LMC ($\sim$11 degrees on the plane of the sky, \citealt{1992AJ....103.1166W}). Note that only M4 is close enough to currently provide a reliable distance measurement directly from parallax. The parallax distance to M4 reported by \citet{Vasiliev_2021} yields a distance modulus of $11.26 \pm 0.04$ mag after accounting for the individual stellar parallax corrections of \cite{EDR3bias} and taking into account spatially correlated systematic errors. In comparison, the EDR3 anchored \rrl{} measurements all provide smaller distance moduli than that from astrometry, with our average M4 measurement 2$\sigma$ smaller. 

Relative discrepancies between individual calibrations can partially be attributed to different assumptions in differential reddening (important with clusters such as M4) and small differences in photometric bands (such as those between IRAC and WISE). Futhermore, even in the Gaia EDR3 release, parallax measurements are still prone to bias with color and magnitude-dependent zero-point offset. \citet{EDR3bias}, as applied, provides a first-order correction to this issue; however, the \rrl{} stars used in our PLZ/PWZ relations possess a quite different color and magnitude than the very bright giant stars often used for direct parallax measurements of globular clusters. 
Note, however, that \citet{Vasiliev_2021} suggests \citet{EDR3bias} might be overcorrecting the Gaia EDR3 parallaxes by $\varpi \sim0.007$ mas, as applied in \citet{2021MNRAS.505.5957B} yielding a slightly larger distance modulus.  It will be interesting to see how future improvements to parallax corrections affect the different types of distance estimates.

\begin{figure*}
\centering
\includegraphics[scale=0.5]{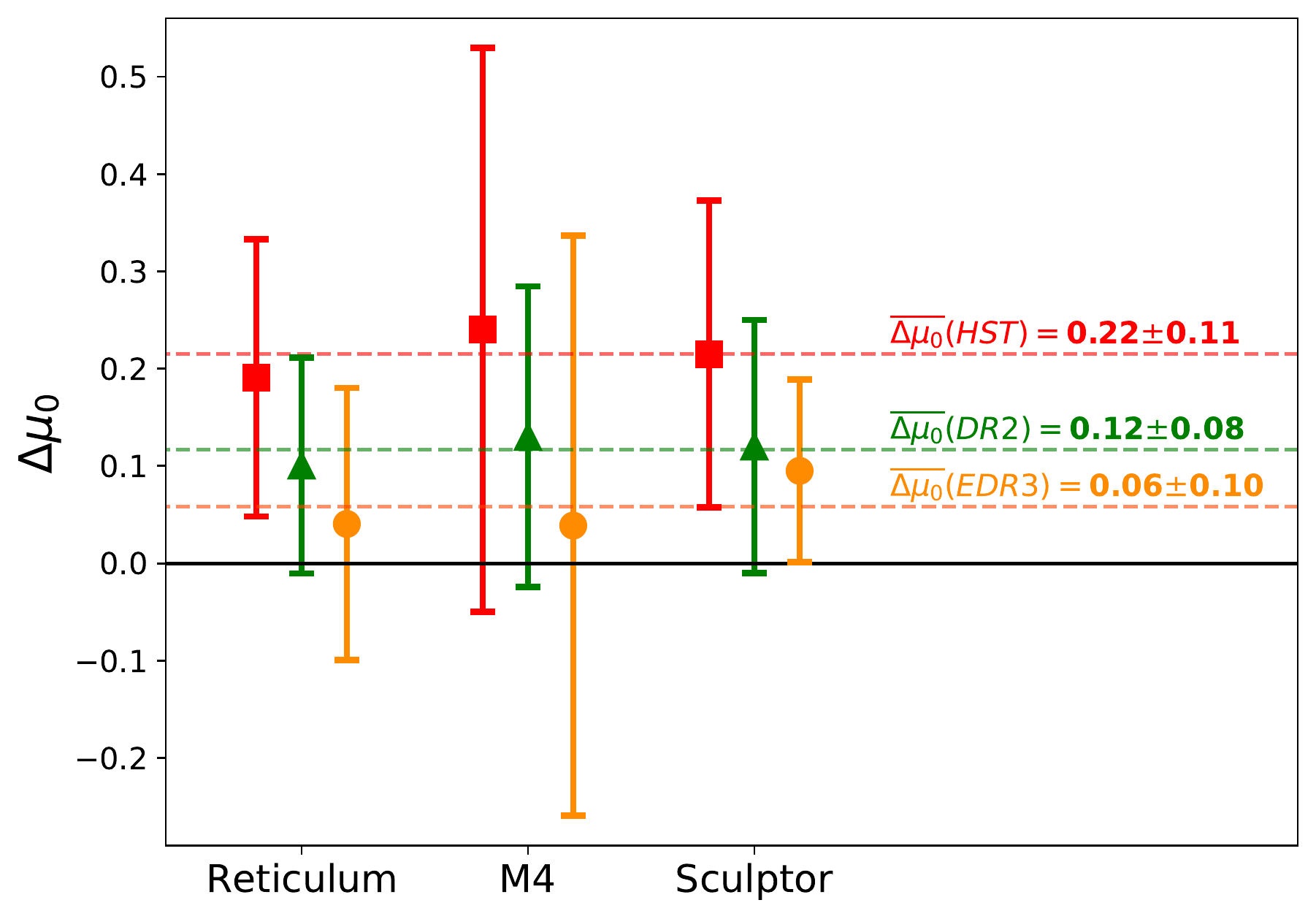}
\caption{The difference in distance moduli between the average literature measurement of those marked in Figure~\ref{fig:moduli} as either HST calibration (red squares), DR2 calib. (green triangles), or EDR3 calib. (yellow circles) and that of this work for three stellar systems: M4, Reticulum, and Sculptor. Errors are propagated from the error in the average distance moduli of this work added in quadrature with the average error of the literature moduli for each calibration type. Horizontal lines represent the average relative distance moduli across all three systems for each type of calibrator used.
 \label{fig:relative_moduli}}
 \end{figure*}

In the mean time, this work corroborates the growing trend of EDR3 anchored \rrl{} distance measurements yielding smaller distance moduli, as shown in works such as \citet{2022MNRAS.513..788G}, \citet{2021MNRAS.503.4719G}, \citet{Nagarajan}, and \citet{2021ApJ...909..200B}. In Figure~\ref{fig:relative_moduli}, we analyze the relative difference between the average distance moduli for M4, Reticulum, and Sculptor calculated in this work and three categories of calibrators: those marked in Figure~\ref{fig:moduli} as \textit{DR2 calib.} \citep{2018MNRAS.480.4138M,2019MNRAS.490.4254N,2022arXiv220508548T,10.1093/mnras/sty2222}, \textit{HST calib.}\citep{2018MNRAS.480.4138M,2013AJ....145..160K,2015ApJ...808...11N,10.1093/mnras/sty2222}, or \textit{EDR3 calib.} from other works \citep{2021MNRAS.503.4719G,2022MNRAS.513..788G,2021ApJ...909..200B,Nagarajan}. For all stellar systems, we note a decreasing offset from the HST calibrations ($\overline{\Delta \mu_0}=0.22\pm 0.11$ mag) to DR2 calibrations ($\overline{\Delta \mu_0}=0.12\pm 0.08$ mag) and continuing to other EDR3 calibrations ($\overline{\Delta \mu_0}=0.06\pm0.10$ mag, consistent with our work). We regard the offset in distance moduli as a difference with the zero-point between the different calibrators. By doing this check, we also show that the systematically smaller EDR3 distances appear to be unrelated to the choice of metallicity term in individual PLZ calibrations, as we see the same trend across all three systems. 

\newpage
\section{Conclusions} \label{sec:conclusion}
In this work, we provide new empirical PLZ/PWZ relations using infrared ($W1$ and $W2$ band) and optical ($V$ band) photometry based on the latest Gaia EDR3 parallaxes. Our relations are calibrated using an entirely field RRL-based sample of both RRab and fundamentalized RRc variables for which homogeneous spectroscopic abundances are available and cover a broad range of metallicities ($-2.5 \la \textrm{[Fe/H]}\la 0.0$) derived from HR spectra and the $\Delta S$ method, using techniques developed by \citetalias{2021ApJ...908...20C}. We derive the period and mean magnitudes used in our calibration directly from the densely populated/ long temporal baselines of the ASASSN and WISE surveys, ensuring a uniform calibration sample. Furthermore, our work utilized the homogeneous processing and quality criterion of Mullen+2021 and Mullen+2022 to uniformly derive the period and mean magnitude, thereby ensuring homogeneity through all of our calibrating variables and a clean sample of RRL stars in ASSASN and WISE surveys. Our overall sample contained $\sim$1000 stars with W1 band magnitudes and $\sim$400 stars in W2, from which we applied Markov chain Monte Carlo to derive our PLZ/PWZ relations.

In order to test the performance of our newly derived PLZ and PWZ relations, we determine distance moduli to the Sculptor dwarf spheroidal galaxy (finding $\bar{\mu}_{0}=19.47 \pm 0.06$) and the globular clusters M4 ($\bar{\mu}_{0}=11.16 \pm 0.05$) and Reticulum ($\bar{\mu}_{0}=18.23 \pm 0.06$). The distance moduli determined through all of our relations are internally self-consistent (within $\lesssim$ 0.05 mag). A comparison with previous literature measurements taken from a variety of methods/anchors reveals that our distance moduli are systematically smaller (by up to $\sim$ 2-3$\sigma$). Compared to direct EDR3 parallax measurements, our \rrl{} EDR3-based PLZ/PWZ distance to M4 has a $\sim$0.1 mag smaller distance modulus. Similarly, other \rrl{} EDR3 anchored calibrations all likewise show to varying extents a systematically smaller distance modulus for the RRL relations when compared to EDR3 parallaxes. We analyze the difference between the average distance moduli for HST, Gaia DR2, and Gaia DR3 calibrations for each of our stellar systems and note across each system a consistent offset relative to this work of $\overline{\Delta \mu_0} (HST)=0.22\pm 0.11$ mag and $\overline{\Delta \mu_0}(DR2)=0.12\pm 0.08$ mag. This suggests that the differences in distance moduli noted are not due to any selection bias in the metallicity term of the PLZ and point towards biases in distance moduli based upon what is used as the zero-point in the PLZ calibrations.


\section{Acknowledgments}
\begin{acknowledgments}
This publication makes use of data products from WISE, which is a joint project of the University of California, Los Angeles, and the Jet Propulsion Laboratory (JPL)/California Institute of Technology (Caltech), funded by the National Aeronautics and Space Administration (NASA), and from NEOWISE, which is a JPL/Caltech project funded by NASA’s Planetary Science Division. 

This publication also makes use of data products from the ASAS-SN project, which has their telescopes hosted by Las Cumbres Observatory. ASAS-SN is supported by the Gordon and Betty Moore Foundation through grant GBMF5490 and the NSF by grants AST-151592 and AST-1908570. Development of ASAS-SN has been supported by Peking University, Mt. Cuba Astronomical Foundation, Ohio State University Center for Cosmology and AstroParticle Physics, the Chinese Academy of Sciences South America Center for Astronomy (CASSACA), the Villum Foundation, and George Skestos.

This work has made use of data from the European Space Agency (ESA) mission
{\it Gaia} (\url{https://www.cosmos.esa.int/gaia}), processed by the {\it Gaia}
Data Processing and Analysis Consortium (DPAC,
\url{https://www.cosmos.esa.int/web/gaia/dpac/consortium}). Funding for the DPAC
has been provided by national institutions, in particular the institutions
participating in the {\it Gaia} Multilateral Agreement.

Guoshoujing Telescope (the Large Sky Area Multi-Object Fiber Spectroscopic Telescope LAMOST) is a National Major Scientific Project built by the Chinese Academy of Sciences. Funding for the project has been provided by the National Development and Reform Commission. LAMOST is operated and managed by the National Astronomical Observatories, Chinese Academy of Sciences.

    Funding for the SDSS and SDSS-II has been provided by the Alfred P. Sloan Foundation, the Participating Institutions, the National Science Foundation, the U.S. Department of Energy, the National Aeronautics and Space Administration, the Japanese Monbukagakusho, the Max Planck Society, and the Higher Education Funding Council for England. The SDSS Web Site is http://www.sdss.org/. The SDSS is managed by the Astrophysical Research Consortium for the Participating Institutions. The Participating Institutions are the American Museum of Natural History, Astrophysical Institute Potsdam, University of Basel, University of Cambridge, Case Western Reserve University, University of Chicago, Drexel University, Fermilab, the Institute for Advanced Study, the Japan Participation Group, Johns Hopkins University, the Joint Institute for Nuclear Astrophysics, the Kavli Institute for Particle Astrophysics and Cosmology, the Korean Scientist Group, the Chinese Academy of Sciences (LAMOST), Los Alamos National Laboratory, the Max-Planck-Institute for Astronomy (MPIA), the Max-Planck-Institute for Astrophysics (MPA), New Mexico State University, Ohio State University, University of Pittsburgh, University of Portsmouth, Princeton University, the United States Naval Observatory, and the University of Washington.

J. P. Mullen and M. Marengo are supported by the National Science Foundation under Grant No. AST-1714534. C. E. Mart\'inez-V\'azquez is supported by the international Gemini Observatory, a program of NSF’s NOIRLab, which is managed by the Association of Universities for Research in Astronomy (AURA) under a cooperative agreement with the National Science Foundation, on behalf of the Gemini partnership of Argentina, Brazil, Canada, Chile, the Republic of Korea, and the United States of America. G. Bono acknowledges partial support from PNRR, CN1, spoke 3.

M. Monelli acknowledges financial support from the ACIISI, Consejer\'ia de 
Econom\'ia, Conocimiento y Empleo del Gobierno de Canarias and the 
European Regional Development Fund (ERDF) under the grant with reference 
ProID2021010075. M. Monelli, also, acknowledges support from the Agencia Estatal de Investigación del 
Ministerio de Ciencia e Innovación (AEI-MCINN) under grant "At the 
forefront of Galactic Archaeology: evolution of the luminous and dark 
matter components of the Milky Way and Local Group dwarf galaxies in the 
Gaia era"  with reference PID2020-118778GB-I00/10.13039/501100011033. Further financial support for M. Monelli comes from the Spanish Ministry of Science 
and Innovation (MICINN) through the Spanish State Research Agency, under 
Severo Ochoa Programe 2020-2023 (CEX2019-000920-S).

\end{acknowledgments}

%

\vspace{5mm}
\facilities{WISE, ASAS-SN, \textit{Gaia}, LAMOST, SDSS-SEGUE}


\software{Astropy \citep{2013A&A...558A..33A}, SciPy \citep{2020SciPy-NMeth}}




\bibliographystyle{aasjournal}



\end{document}